\documentclass[10pt]{article}

\usepackage[utf8]{inputenc}
\usepackage[T1]{fontenc}
\usepackage[square,sort&compress]{natbib}    
\usepackage{booktabs}
\usepackage[french,english]{babel}
\usepackage[top=1in,left=1.5in,bottom=1in,right=1in]{geometry}

\usepackage{relsize}
\usepackage{amsmath, amssymb}
\usepackage{graphicx}
\usepackage{subfig}
\usepackage{authblk}
\usepackage{eurosym}

\usepackage[
  bookmarks       = true,             
  unicode         = true,             
  pdftoolbar      = true,             
  pdfmenubar      = true,             
  pdffitwindow    = false,            
  pdfstartview    = {FitH},           
  pdfnewwindow    = true,             
  colorlinks      = true,             
  linkcolor       = blue,             
  citecolor       = blue,            
  filecolor       = cyan,             
  urlcolor        = magenta           
]{hyperref}
\usepackage{algorithm}
\usepackage{algorithmicx}
\usepackage{listings}
\usepackage{algcompatible}
\usepackage{algpseudocode}
\usepackage{tikz}
\usepackage{calc}
\usepackage{pgfplots}
\usepackage{pdfpages}
\pgfplotsset{compat=1.14}
\usepackage{tikzducks}
\usetikzlibrary{snakes}
\usetikzlibrary{positioning}
\usetikzlibrary{shapes,arrows}
\usetikzlibrary{automata,arrows,positioning,calc}

\makeatletter
\newcounter{algorithmicH}
\let\oldalgorithmic\algorithmic
\renewcommand{\algorithmic}{%
  \stepcounter{algorithmicH}
  \oldalgorithmic}
\renewcommand{\theHALG@line}{ALG@line.\thealgorithmicH.\arabic{ALG@line}}
\makeatother

\usepackage{kpfonts}
\usepackage[protrusion = true, expansion = true]{microtype} 
\usepackage{multicol}
\usepackage{caption}            
\usepackage{siunitx}            
\usepackage{graphicx}           
\usepackage{fancyhdr}           
\usepackage{amsthm}
\usepackage[capitalise,nameinlink,noabbrev]{cleveref}
\crefname{equation}{}{}
\crefname{enumi}{}{}

\newtheorem{prop}{Proposition}

\newtheorem{remark}{Remark}
\newtheorem{ex}{Example}

\usepackage{xspace}

\newcommand*{\Prob}{\mathbb{P}}
\newcommand*{\E}{\mathbb{E}}

\newcommand*{\e}{\mathrm{e}}

\begin{document}

\title{Hashpower allocation in Pay-per-Share blockchain mining pools}
\author[1]{Hansjoerg Albrecher\footnote{Email: \href{mailto:hansjoerg.albrecher@unil.ch}{hansjoerg.albrecher@unil.ch}.}}
\author[2]{Jean-Pierre Fouque\footnote{Email: \href{mailto:fouque@pstat.ucsb.edu}{fouque@pstat.ucsb.edu}.}}
\author[3]{Pierre-O. Goffard\footnote{Email: \href{mailto:goffard@unistra.fr}{goffard@unistra.fr}.}.\footnotesize}
\affil[1]{Department of Actuarial Science, Faculty of Business and Economics (HEC), University of Lausanne and Swiss Finance Institute, UNIL-Chamberonne, CH-1015 Lausanne}
\affil[2]{Department of Statistics and Applied Probability, University of California Santa Barbara, CA 93106-3110, USA}
\affil[3]{Université de Strasbourg, Institut de Recherche Mathématique Avancée, Strasbourg, France}

\date{\today }

\maketitle
\vspace{3mm}

\begin{abstract}
{
Mining blocks in a blockchain using the \textit{Proof-of-Work} consensus protocol involves significant risk, as network participants face continuous operational costs while earning infrequent capital gains upon successfully mining a block. A common risk mitigation strategy is to join a mining pool, which combines the computing resources of multiple miners to provide a more stable income. This article examines a Pay-per-Share (PPS) reward system, where the pool manager can adjust both the share difficulty and the management fee. Using a simplified wealth model for miners, we explore how miners should allocate their computing resources among different mining pools, considering the trade-off between risk transfer to the manager and management fees.
}
\end{abstract}
{\it Keywords:} Optimal dividend; mean-variance trade-off; blockchain; mining pools.\\

\section{Introduction}
A blockchain is a distributed ledger maintained through consensus in a peer-to-peer network. In a permissionless blockchain, anyone can join the network, allowing it to grow indefinitely. Achieving consensus in a large, decentralized network is a technical challenge, requiring a protocol that enables nodes to agree on a common transaction history. In the \textit{Proof-of-Work} (PoW) protocol, nodes—known as miners—compete to solve a cryptographic puzzle using brute-force search algorithms. The first miner to find a valid solution appends a new block to the chain. Mining is computationally intensive and consumes significant amounts of electricity, as miners operate their machines continuously. This operational cost is offset by rewards distributed in the blockchain’s native cryptocurrency. This incentive mechanism underpins the Bitcoin protocol of \citet{Na08} and other major public, permissionless blockchain platforms.\\

\noindent The wealth of a miner can be modeled as a stochastic process:  
\begin{equation}\label{eq:miner_solo_wealth_process}  
X_t = x - c\cdot t + b\cdot N_t, \quad t\geq0,  
\end{equation}  
where \( x > 0 \) represents the initial capital, \( c \) is the operational cost, \( b \) is the block discovery reward, and \( (N_t)_{t\geq0} \) is a Poisson process with intensity \( \lambda \). The variance of this process exposes miners to the risk of insolvency. The model in \eqref{eq:miner_solo_wealth_process} was introduced by \citet{Hansjoerg2022} to assess such risks. The model has been extended in \citet{Li2024} with a mean-field game setup to discuss whether a miner should mine at full capacity. To mitigate income volatility, miners form mining pools, combining their computational resources to increase the frequency of capital gains and achieve more stable earnings. A pool manager coordinates this collective effort by evaluating each participant's contribution and distributing rewards accordingly. This paper focuses on a widely used reward distribution system known as \textit{Pay-per-Share} (PPS). In this scheme, miners submit \textit{shares} to the pool manager as proof of their work. A \textit{share} is a solution to a cryptographic puzzle with a lower difficulty threshold than that required for block mining. In the \textit{Pay-per-Share} reward system, the pool manager offers a reward \( \tilde{b} < b \) to miners for each submitted \textit{share}. The wealth of a miner then follows the stochastic process
\begin{equation}\label{eq:miner_inpool_wealth_process}
\widetilde{X}_t = x - c\cdot t + \tilde{b} \cdot \tilde{N}_t, \quad t\geq0,
\end{equation}
where \( (\tilde{N}_t)_{t\geq0} \) is a Poisson process with intensity \( \tilde{\lambda} > \lambda \), which exceeds that of \( (N_t)_{t\geq0} \) in \eqref{eq:miner_solo_wealth_process}. The \textit{Pay-per-Share} reward system has been shown to be both fair and incentive-compatible in the work of \citet{Schrijvers2017}. For an overview of mining pools and reward distribution schemes, the reader is referred to the early survey by \citet{rosenfeld2011analysis}, where the Pay-per-Share system is presented with a fixed relative difficulty for finding a share. One advantage of the \textit{Pay-per-Share} system is that it transfers risk from miners to pool managers. Models \eqref{eq:miner_solo_wealth_process} and \eqref{eq:miner_inpool_wealth_process} can be used to quantify the risk miners undertake, particularly through ruin probabilities, as demonstrated in \citet{Albrecher2022}. These models closely resemble those used in insurance risk management; for a broader discussion, see the textbook by \citet{Asmussen_2010}. Our focus in this work is on the potential income of miners rather than their probability of failure. Miners evaluate various offers from different mining pools and must decide how to allocate their computing resources. Our goal is to provide insights into how mining pool formation impacts the decentralization of a \textit{Proof-of-Work} blockchain. If mining resources become concentrated within a few large mining pools, this raises concerns about the blockchain’s decentralized nature, see for instance \citet{Cong2020} .
 \\

\noindent We approach this problem by defining various objective functions that miners might seek to maximize. We begin with a simple mean-variance trade-off, assuming that mining within a pool is always less profitable on average but provides a more stable income. Next, we examine an objective function inspired by the classical dividend control problem introduced by \citet{de1957impostazione}. Concretely, the value of the mining portfolio (that is, the partition between the participation in mining pools and solo mining) for a miner is measured in terms of the expected discounted cashflows (dividend payments) out of that portfolio until it is depleted, when these dividend payments are executed in the best possible way. The topic of optimal dividend payment strategies has been extensively studied in the insurance risk literature; for comprehensive surveys, see \citet{Avanzi2009} and \citet{Albrecher2009}. The surplus processes of miners, as given in \eqref{eq:miner_solo_wealth_process} and \eqref{eq:miner_inpool_wealth_process}, correspond exactly to the dual risk model with constant capital gains. It is well known that a barrier strategy is optimal in this setting; see \citet{Avanzi2007}, \citet{Avanzi2008}, and \citet{Bayraktar2013}. Since the processes defined in \eqref{eq:miner_solo_wealth_process} and \eqref{eq:miner_inpool_wealth_process} are Lévy processes, we leverage fluctuation theory to develop numerical routines for computing the optimal barrier level and the associated value function. This requires studying the scale functions of the underlying Lévy processes. For an overview of fluctuation theory and scale functions, we refer the reader to the book by \citet{Kyprianou2014}. We provide a way to understand how a network of miners distribute their hashpower across a set of mining pools offering various deals in terms of pool management fee and difficulty of finding a \textit{share}.
 \\

\noindent The remainder of the paper is organized as follows.  \cref{sec:blockchain_mining_pool} describes the mining process and the formation of mining pools. We further define several objective functions that miners could use to allocate their hashpower across multiple mining pools. We consider a mean-variance trade-off in \cref{sec:mean_variance} and use the optimal expected discounted dividends associated to the miner's wealth process in \cref{sec:optimal_revenue}. A numerical illustration is provided in  \cref{sec:numerical_illustrations}, which involves assessing the decentralization in simulated blockchain networks.
\section{Blockchain and mining pool}\label{sec:blockchain_mining_pool}
\subsection{Blockchain mining process}\label{ssec:blockchain_mining}
A block consists of a header and a list of ``transactions'' that represent the information recorded on the blockchain. The header typically includes the creation date and time of the block, the block height (which is the index within the blockchain), the hash of the block, and the hash of the previous block. The hash of a block is obtained by concatenating the header and the transactions into a large character string, forming a ``message,'' to which a hash function is applied. A hash function maps data of arbitrary size to fixed-sized values. In blockchain applications, these hash functions must be cryptographic, meaning they are quick to compute, one-way, and deterministic. It must be nearly infeasible to generate a message with a given hash value or to find two messages with the same hash value. Even a small change in the message should dramatically alter the hash value, making the new hash appear uncorrelated with the previous one. We will not delve into the construction of such cryptographic hash functions; interested readers are referred to the work of \citet{AlKuwari2011}. In the Bitcoin blockchain and many other applications, the standard is the SHA-256 function, which converts any message into a 256-bit hash value. This value is usually translated into a hexadecimal digest. For instance, the hash value of the title of this manuscript is:

$$
\texttt{98b1146926548f6b57c4347457713ff2f035beda9c93f12fbc9b202e9c512e80}.
$$

\begin{figure}[!ht]
    \includegraphics[width = \textwidth]{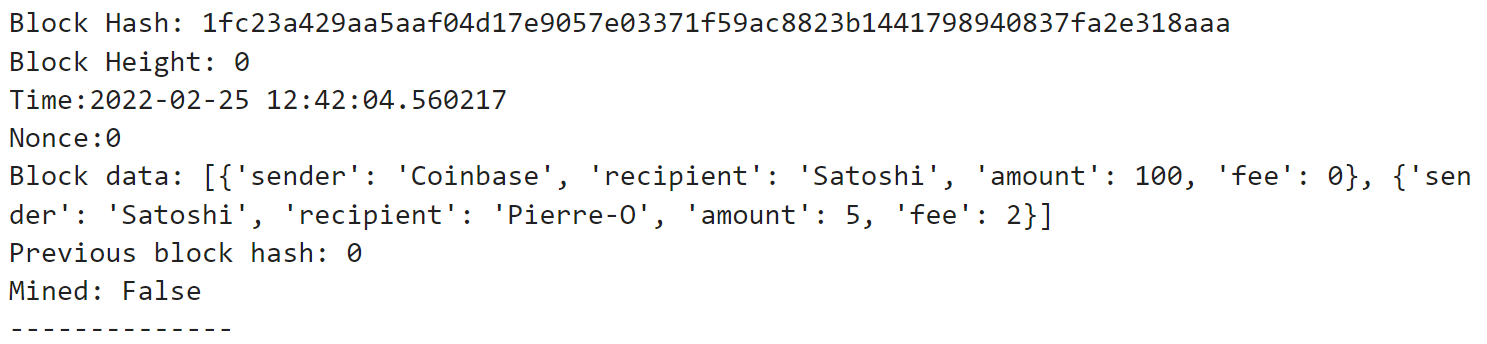}
    \captionsetup{width=0.8\textwidth}
    \centering
    \caption{A block that has not been mined yet.}
    \label{fig:block_not_mined}
\end{figure}

The information recorded in a public blockchain can be retrieved by anyone and accessed through a blockchain explorer such as \href{https://www.blockchain.com/}{blockchain.com}. For example, the content of the block at height \#724724 can be viewed through the following link: \href{https://www.blockchain.com/btc/block/0000000000000000000954d42e8ced7017448cb9f39b364e371a1eec6e34463b}{block content}. Mining a block involves finding a block hash value lower than a specified target, which can only be achieved through brute force search due to the properties of cryptographic hash functions. In practice, the search for an appropriate hash value, referred to as a solution, is conducted by appending a nonce to the block message before applying the hash function. A nonce is a 32-bit number drawn at random by miners until a nonce that results in a proper block hash value is found. For illustration, consider the block in \cref{fig:block_not_mined}.

\begin{figure}[!ht]
    \includegraphics[width = \textwidth]{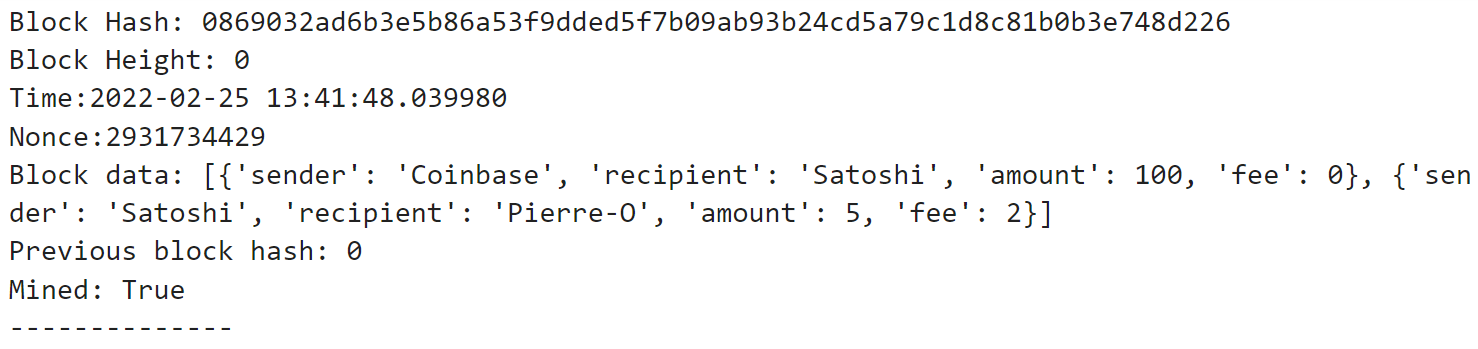}
    \captionsetup{width=0.8\textwidth}
    \centering
    \caption{A mined block with a hash value having one leading zero.}
    \label{fig:block_mined}
\end{figure}

\noindent The hash value in decimal notation is $1.43 \times 10^{76}$, while the maximum value for a 256-bit number is $2^{256} - 1 \approx 1.16 \times 10^{77}$. We refer to the latter as the maximal target and denote it by $T_{\max}$. The Proof-of-Work protocol sets a target $T < T_{\max}$ and requires miners to find a nonce such that the hash value of the block is smaller than $T$. Practitioners often refer to the \textit{difficulty}, which is defined as $D = T_{\max} / T$. If the difficulty is one, any hash value is acceptable. Increasing the difficulty reduces the set of allowable hash values, making the problem harder to solve. A hash value is then called \textit{acceptable} if its hexadecimal digest starts with a given number of zeros. If we set the difficulty to $2^4$, then the hexadecimal digest of the block's hash must start with at least one leading zero. This makes the hash value of the block in Figure \ref{fig:block_not_mined} not acceptable. After completing the nonce search, we obtain the block in Figure \ref{fig:block_mined}.

\noindent Note that it took $5$ attempts to find this nonce. The number of needed trials is geometrically distributed with parameter $1 / D$, which means that with a difficulty of $D = 2^4$ it takes on average $16$ trials. The bitcoin protocol adjusts the difficulty automatically every $2,016$ block discoveries so as to (globally) maintain one block discovery every $10$ minutes on average. The time between two block discoveries depends on the number of hash values computed by the network at a given instant. At the time of writing, the network computes $182.58$ Exahashes per second (EH/s) and the difficulty is $27,967,152,532,434$.\footnote{Source: \href{https://www.bitcoinblockhalf.com/}{bitcoinblockhalf.com}} For an exhaustive overview of the mining process in the bitcoin blockchain, we refer the reader to the book of Antonopoulos \cite[Chapter 12]{Antonopoulos2017}. As each trial (of the system) for mining a block is independent of the others and leads to a success with very small probability, the overall number of successes is binomially distributed and will be very well approximated by a Poisson random variable. This justifies the Poisson process assumption made in the sequel to model the block arrival and the reward collecting processes. Empirical studies of the block inter-arrival times data tend to confirm this hypothesis, see the work of \citet{Bowden2020}.\\

These considerations validate the model \eqref{eq:miner_solo_wealth_process} for a miner's wealth over time, as outlined in the introduction. The wealth of a miner is given by
\[
  X_t = x - c \cdot t + N_t \cdot b, \quad t \geq 0,
\]
where \( x \) is the initial wealth, \( c \) is the operational cost, \( (N_t)_{t \geq 0} \) is a Poisson process with intensity \( \lambda > 0 \), and \( b \) is the reward for finding a block through successful hashing. The intensity \( \lambda \) of the Poisson process is related to the miner's hashrate \( h \) (expressed in hashes per second) and the current difficulty \( D \) of finding a block. Specifically, \( \lambda = h / D \), representing the number of blocks found per unit of time. The miner's hashrate incurs an operational cost \( c \). To compute hash values efficiently, miners must invest in specialized equipment called Application Specific Integrated Chips (ASICs). These devices require significant investment and space for storage, along with cooling, ventilation, and noise isolation systems. Additionally, technicians, engineers, and security personnel are needed to maintain operations and ensure safety. Data storage units are also necessary for logging data and monitoring performance. The facility's electricity consumption, measured in Watts, is substantial. The operational cost \( c \) is derived from the sum of these elements.
The reward \( b \) for finding a new block includes the bounty for the block and the transaction fees. 
\begin{remark}\label{rem:transaction_fee}
The bounty for finding a new block in the Bitcoin blockchain allows us to neglect transaction fees as a first approximation. This approximation is valid at the time of writing, as the block reward of \(\text{BTC } 3.125\) is quite substantial. However, this may change in the future, as the block reward is halved approximately every four years\footnote{A countdown until the next block reward halving is provided on \url{https://www.bitcoinblockhalf.com/}}. When accounting for transaction fees, one must replace the fixed \(b\) with a stochastic process in the wealth process \eqref{eq:miner_solo_wealth_process} of the miner. The use of an independent and identically distributed (i.i.d.) sequence was considered in \citet{Albrecher2022}, and the introduction of time dependency was explored in \citet{Albrecher2024}. Proper statistical studies of the dynamics of transaction fees are readily available in the literature; see, for instance, \citet{Moeser2015} and \citet{Easley2019}.

\end{remark}
Bitcoin users include fees in their transactions to incentivize miners to process them. The total fees in a block are added to the block's bounty. Miners must consider the exchange rate between the blockchain's native cryptocurrency and fiat currencies like USD or EUR. Before starting a mining operation, miners must ensure it is profitable on average, which requires satisfying the net profit condition: \( \lambda \cdot b > c \). In \cref{ex:mining_farm}, we aim to provide an estimate of the costs associated with mining the Bitcoin blockchain in France.

\begin{ex}\label{ex:mining_farm}
At the time of writing, a high-quality ASIC can compute 53 Terahashes (\(10^{12}\) hashes) per second (TH/s) and consumes 3 kW of electricity. The entire Bitcoin network currently computes approximately 805.32 Exa hashes (\(10^{18}\) hashes) per second. A new miner aiming to control 0.005\% of the network's hashpower would need to acquire 800 ASICs that cost approximately $\euro1,000$, this is already an initial investment of $\euro800,000$. The monthly electricity consumption for these ASICs would be 1,728,000 kWh. In France, the electricity rate ranges from €0.15 to €0.20 per kWh. The ASICs must be stored in racks with adequate spacing, and the facility must include cooling and ventilation systems, as well as office space for workers. A team of 8 to 15 people, including technicians, engineers, administrative staff, and security personnel, is required. The building should be approximately 360 square meters in size.

The monthly operational cost is divided as follows:

\begin{equation}
c = \underbrace{276,480}_{\text{electricity}} + \underbrace{4,000}_{\text{rent}} + \underbrace{40,000}_{\text{wages}} = 320,480.
\end{equation}

On the revenue side, the block reward is currently $3.125$ bitcoins per block, with an average of $0.04$ bitcoin added from transaction fees. The network mines approximately $4,320$ blocks per month, and $0.005\%$ of these are discovered by our miner, amounting to about 0.22 blocks per month. At the current exchange rate of $\euro75,007$ per bitcoin, the monthly revenue is:

\begin{equation}
\lambda \cdot b = 0.22 \times 3.165 \times 75,007 = 54,014.
\end{equation}

Given these figures, it appears that mining Bitcoin in France is unlikely to be profitable.

\end{ex}
\begin{remark}
Many sources of randomness could be considered to refine our simple model, such as the volatility of the exchange rate between cryptocurrency and fiat currency, fluctuations in electricity prices, and variations in transaction fees included in the block reward. For the sake of simplicity, we have made certain assumptions to facilitate understanding and to derive tractable formulas suitable for numerical evaluation.
\end{remark}
\cref{ex:mining_farm} demonstrates that even with a substantial initial investment, the rate of block discovery is low, resulting in infrequent capital gains. This could deter potential investors from supporting such ventures. Consequently, the blockchain ecosystem has seen the rapid emergence of mining pools, where miners combine their resources to achieve more stable mining revenues. The next subsection provides an overview on how mining pools operate.

\subsection{Mining pools}\label{ssec:mining_pool}
A mining pool is a collaborative group of miners who combine their computational resources over a network to enhance their chances of successfully mining a block. Miners join these pools to achieve more consistent and predictable earnings, as the combined effort increases the likelihood of solving the complex mathematical problems required to add a new block to the blockchain. By participating in a pool, miners can receive a steady income stream, reducing the variance and unpredictability associated with solo mining.\\

\noindent When joining mining pools, miners demonstrate their work to pool managers by submitting \textit{shares}. \textit{Shares} are hash values that meet a lower difficulty level than the actual cryptopuzzle. The number of \textit{shares} submitted to the pool manager determines how the block reward, collected by the pool, is distributed among the miners. The reward redistribution system of the mining pool is a key feature of the mining pool, we present the main approaches hereafter.\footnote{A list of the mining pools is maintained on the webpage \url{https://miningpoolstats.stream/} for each cryptocurrency. We take the acronym below from this source which also provides a snap shot of the hashpower distribution of the mining pools.}

\paragraph*{Proportional system (PROP)} The simple proportional reward system, used in the early days of mining pools, distributes rewards to participants based on the number of shares they submit since the last block discovery. However, this system has several weaknesses:

\begin{enumerate}
    \item Payments only occur when a block is found by a pool member, causing the value of submitted \textit{shares} to decrease as the time between block discoveries increases. This can lead to pool-hopping behavior, where miners switch to pools with shorter block discovery times.
    \item Miners could withhold the submission of a \textit{share} that solves the block puzzle to manipulate their share of the pool's hashpower.
    \item There is little to no risk transfer from miners to pool managers.
\end{enumerate}
\paragraph*{Pay-Per-Last-N-Shares (PPLNS)} This reward system resembles the proportional reward system, except that the contribution of the participants is evaluated only over the last \(N\) shares prior to finding a block. This system circumvents the first problem of the proportional system. Again, no risk is transferred to the pool manager, which limits her probability of failure to low or non-existent, as shown, for instance, in \citet{Can2022}.
 
\paragraph*{Pay-Per-Share (PPS)} Miners receive a payment each time they submit a share. Since shares are easier to find than blocks, the reward per share is smaller, but the frequency of rewards is higher. It is important to note that the PPS system addresses all the issues of the proportional (PROP) system. The value of a share is determined by the difficulty of finding a \textit{share} compared to finding a block, without accounting for transaction fees, as these are unknown when a share payment is due. Two variants of the PPS scheme are used to include transaction fees in the miner's income. The first is called Full Pay-Per-Share (FPPS), where the unknown transaction fee is replaced by a proxy equal to the average of the transaction fees collected in recent blocks. The second variant is called Pay-Per-Share Plus (PPS+), where transaction fees are redistributed whenever a block is found, following the Pay-Per-Last-N-Shares (PPLNS) scheme.
  
\paragraph*{Solo Mining Pool (SOLO)} There exist mining pools in which miners engage in solo mining, thereby receiving rewards only for the blocks they find. One might question the purpose of joining such a pool, as joining a pool typically involves paying a management fee. We outline the benefits of joining a mining pool, beyond income stability, in \cref{rem:mining_pool_advantage}.

\paragraph*{Peer-To-Peer Mining Pool (P2POOL)} A P2POOL mining pool maintains its own blockchain to keep track of the shares submitted by the participants. The participants act as nodes in this secondary blockchain, and a block associated with a share that is a valid solution is automatically broadcast to the primary blockchain. Such mining pools do not have a pool manager, and the distribution of rewards is conducted using either a Proportional (PROP) or Pay-Per-Last-N-Shares (PPLNS) scheme.

\begin{remark}\label{rem:mining_pool_advantage}
Mining pools provide several benefits beyond stable income. These benefits include cloud storage of blockchain data, as miners need to maintain and update their view of the blockchain, which can be demanding in terms of data storage and bandwidth. Mining pools typically have better connectivity to the blockchain network, ensuring faster propagation of blocks and reducing the likelihood of stale blocks, which can occur when blocks are mined but not relayed quickly enough to the network. Mining pools often provide comprehensive monitoring dashboards that display metrics such as network hash rate, cryptocurrency prices, and mining returns. These metrics enable miners to make informed decisions and optimize their strategies. These services justify the fact that miners are charged a fee even in Proportional (PROP), Pay-Per-Last-N-Shares (PPLNS), and Solo (SOLO) mining pools, despite the absence of risk transfer. Note that the pool management fees in Pay-Per-Share (PPS) pools are larger than in other types of mining pools.
\end{remark}

We focus in this work on the PPS mining pool which is mainly used by major mining pools. When joining such a pool, the miner's initial wealth and operational costs remain unaffected, and the wealth process for a miner in the pool is given by:

\begin{equation*}
\widetilde{X}_t = x - c \cdot t + \tilde{b} \cdot \tilde{N}_t, \quad t \geq 0,
\end{equation*}
where \(\tilde{b} < b\) and \(\left(\tilde{N}_t\right)_{t \geq 0}\) is a Poisson process with intensity \(\tilde{\lambda} > \lambda\). As risk is transferred from miners to pool managers, managers apply a fee, reducing profitability:
\[
\tilde{\lambda} \cdot \tilde{b} < \lambda \cdot b.
\]
The variance of the wealth process when mining in a pool is lower than when mining solo, as \(\tilde{\lambda} \cdot \tilde{b}^2 < \lambda \cdot b^2\). The parameters \(\tilde{\lambda}\) and \(\tilde{b}\) can be set using a relative difficulty \(\delta\) for finding a share compared to a block, and a pool management fee \(f\), as illustrated in \cref{ex:mining_pool}.

\begin{ex}\label{ex:mining_pool}
The pool manager may set the difficulty of finding a share to be \(\delta\) times lower than finding a block. If a miner's hash rate results in a block discovery rate of \(\lambda\), then her share discovery rate is:
\begin{equation}
\tilde{\lambda} = \frac{\lambda}{\delta}.
\end{equation}
The reward for finding a share is then:
\begin{equation}
\tilde{b} = \delta \cdot b \cdot (1 - f),
\end{equation}
where \(f\) is the manager's fee. A common setting among mining pools is a 2.5\% fee with a difficulty for finding a share equal to \(1/2^{32}\), as seen in the \href{https://academy.braiins.com/en/braiins-pool/rewards-and-payouts/#fpps-specification}{Braiins mining pool}.
\end{ex}
\noindent We study the question of how a miner should allocate her hashpower among several Pay-Per-Share (PPS) mining pools. The problem of hashpower allocation across several Pay-Per-Last-N-Shares (PPLNS) mining pools has been considered in \citet{Chatzigiannis2022}. PPLNS mining pools witness competition among pool participants, which raises interesting game-theoretical questions. Such considerations are absent from the PPS mining pools from the miner's point of view. The focus is more on the risk management aspect from the miner's perspective, hence the choice of a risk-theoretic indicator in \cref{sec:optimal_revenue}.

\subsection{Miner's wealth process}\label{ssec:wealth_process}
The computational capacity of a miner is characterized by a block discovery rate, \(\lambda > 0\), and an operational cost, \(c > 0\). There are \(n \in \mathbb{N}\) mining pools, each offering a share discovery rate, \(\lambda_k\), associated with a reward, \(b_k\), for \(k = 1, \ldots, n\). Our miner retains the option to mine independently; thus, we introduce an artificial mining pool, \(k = 0\), where \(b_0 = b\) and \(\lambda_0 = \lambda\). Without loss of generality, we assume that:
\[
\lambda_0 < \ldots < \lambda_n \text{ and } b_0 > \ldots > b_n.
\]
The miner's computational resources are allocated among the mining pools according to a weight vector:
\[
w = \begin{pmatrix} w_0 & w_1 & \ldots & w_n \end{pmatrix} \in \Delta^{n},
\]
where \(\Delta^{n} = \left\{w \in \mathbb{R}^{n+1}; \sum_{i=0}^n w_i = 1\right\}\) represents the unit simplex. The rate at which rewards are obtained from each mining pool is given by:
\[
\mu_k = \lambda_k w_k, \text{ for } k = 0, \ldots, n.
\]
The \(\mu_k\) values represent the intensities of \(n+1\) independent Poisson processes. Consequently, the arrival of rewards can be modeled by a Poisson process, \((N_t)_{t \geq 0}\), with intensity \(\mu = \sum_{k=0}^n \mu_k\). The probability that a reward originates from mining pool \(k\) is \(p_k = \mu_k / \mu\), with the reward amount being \(b_k\). Assuming the miner starts with an initial capital \(x > 0\), her wealth over time is expressed as:
\begin{equation}\label{eq:surplus_miner_split_accross_mining_pools}
X_t = x - c \cdot t + \sum_{i=1}^{N_t} B_i,
\end{equation}
where the \(B_i\) values form a sequence of independent and identically distributed random variables with the following distribution:
\[
\Prob(B = b_k) = \frac{\mu_k}{\mu} = p_k, \text{ for } k = 0, \ldots, n.
\]
\begin{remark}
Should the miner choose to operate outside the mining pools, \(w_0\) is set to 1, reverting to the wealth process defined in the introduction as \(\eqref{eq:miner_solo_wealth_process}\). Conversely, if the miner focuses solely on mining pool \(k\), \(w_k\) is set to 1, aligning with \(\eqref{eq:miner_inpool_wealth_process}\).
\end{remark}
\noindent We define a value function as follows:
\[
w \in \Delta^n \mapsto V(w) \in \mathbb{R}_+.
\]
The miner's objective is to determine \(w^\ast\), defined by:
\[
w^\ast = \underset{w \in \Delta^n}{\text{argmax}} V(w).
\]
In the subsequent sections, we outline various objective functions which the miner aims to maximize for strategic decision-making. Section~\ref{sec:mean_variance} introduces straightforward criteria based on the mean and variance of the wealth process, while Section~\ref{sec:optimal_revenue} delves into a more sophisticated criterion that accounts for risk through the probability of ruin (i.e., depletion of the surplus process), rather than variance.\\

\section{Mean-variance trade-offs}\label{sec:mean_variance}
The expected wealth of the miner of \cref{ssec:wealth_process} (associated to the wealth process \eqref{eq:surplus_miner_split_accross_mining_pools}) at time \( t \geq 0 \) is given by
\begin{equation}\label{eq:expected_wealth}
\E(X_t) = x - c t + t \sum_{k=0}^n w_k \lambda_k b_k.
\end{equation}
The variance of the wealth of our miner at time \( t \geq 0 \) is given by
\begin{equation}\label{eq:wealth_variance}
\text{Var}(X_t) = t \sum_{k=0}^n w_k \lambda_k b_k^2.
\end{equation}
The objective functions defined hereafter rely on the miner's preferences in terms of risk through an explicit risk aversion parameter \(\gamma\) or a pre-specified level of the variance of the wealth process \(\sigma^2\).

\subsection{Mean-variance optimization accounting for risk aversion}\label{ssec:mean_variance_optimization}
Let \( T \) represent a specified time horizon. We consider the following objective function:
\begin{equation}\label{eq:mean_variance_objective_function}
V(w) = \E(X_T) - \gamma \text{Var}(X_T),
\end{equation}
where \(\gamma > 0\) is a risk aversion parameter that reflects the miner's preferences. By incorporating the expressions for expectation \eqref{eq:expected_wealth} and variance \eqref{eq:wealth_variance} into the value function \eqref{eq:mean_variance_objective_function}, we obtain:
\[
V(w) = x - cT + T \cdot \sum_{k=0}^n w_k \cdot \lambda_k \cdot (b_k - \gamma b_k^2).
\]
From this, we deduce that the optimal strategy is \( w_{k^\ast} = 1 \), where:
\[
k^\ast = \underset{k=0,\ldots, n}{\text{argmax}}\, \lambda_k \cdot (b_k - \gamma b_k^2).
\]
This result is independent of the time horizon \( T \). When using this criterion, the miner's decision depends solely on the risk aversion parameter \(\gamma\). Consequently, miners with identical preferences should all opt for the same mining pool. The influence of the risk aversion parameter is further illustrated in \cref{ex:mean_variance_risk_aversion}.

\begin{ex}\label{ex:mean_variance_risk_aversion}
Let the reward for finding a block be \( b = \text{BTC } 3.125 \). Consider a miner with the following characteristics:
\[
\lambda = 6, \quad c = 14.42.
\]
The miner has the option to mine independently or join a Pay-Per-Share (PPS) mining pool. The mining pools have the following pool management fees and corresponding rewards per share:
\[
\begin{array}{l l l}
\text{Pool Management Fees:} & \lambda_1 = 6.06, & \lambda_2 = 7.06, \quad \lambda_3 = 8, \\
\text{Rewards per Share:} & b_1 = 3.08, & b_2 = 2.62, \quad b_3 = 2.1.
\end{array}
\]When the risk aversion parameter is low, the miner prefers solo mining. Conversely, if the risk aversion parameter is high, the miner opts for Pool \(3\). As \(\gamma\) increases, the miner briefly prefers Pool $1$, then Pool \(2\) and finally Pool \(3\). 
Figure \ref{fig:risk_aversion_mean_variance_tradeoff} illustrates \(\lambda_k \cdot (b_k - \gamma b_k^2)\) as a function of the risk aversion parameter \(\gamma\) and the chosen mining pool \(k = 0, \ldots, 3\).
\begin{figure}[!ht]
\centering
    \includegraphics[width = 0.6\textwidth]{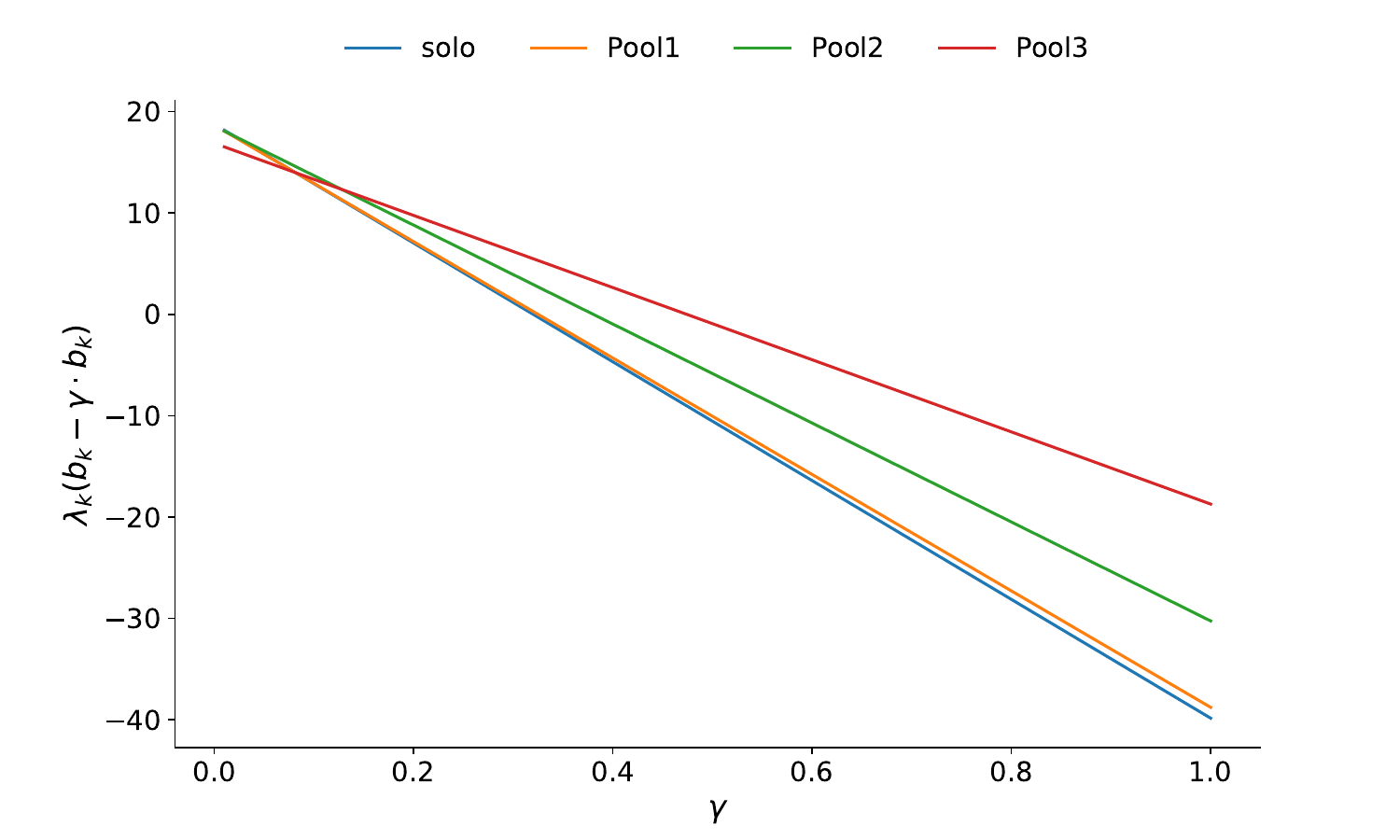}
    
    \caption{Mean-variance tradeoff depending on the risk aversion parameter in the presence of   $n = 3$ pools.}
    \label{fig:risk_aversion_mean_variance_tradeoff}
\end{figure}
\end{ex}
\subsection{Mean-Variance efficient frontier approach}\label{ssec:mean_variance_frontier}
The approach described above involves selecting a risk aversion parameter \(\gamma\) to guide the miner's decision on allocating resources to a specific pool. An alternative method to achieve a mean-variance trade-off is to choose a level of risk \(\sigma^2\) and maximize the expected wealth for that given level of risk. This method defines an efficient frontier, which becomes apparent when plotting investment decisions on an expectation versus variance  plot. The miner must then solve the following constrained optimization problem:
\begin{equation}\label{eq:efficient_frontier}
\underset{w \in \Delta^n}{\max}\, \mathbb{E}(X_T) \text{ such that } \text{Var}(X_T) = \sigma^2.
\end{equation}
Once again, the time horizon \(T\) is irrelevant as it does not affect the solution; hence, we set \(T = 1\). For the problem to have a solution, we must impose the condition \(\lambda_n b_n^2 \leq \sigma^2 \leq \lambda_0 b_0^2\). The optimization problem can be rewritten as
\begin{equation}\label{eq:efficient_frontier_rewritten}
\underset{w \in \Delta^n}{\max}\, \sum_{k=0}^n w_k \lambda_k b_k \text{ such that } \sum_{k=0}^n w_k \lambda_k b_k^2 = \sigma^2,
\end{equation}
where
\[
\lambda_0 b_0 \geq \ldots \geq \lambda_n b_n \text{ and } \lambda_0 b_0^2 \geq \ldots \geq \lambda_n b_n^2.
\]
The mining pools represent the extreme points of a convex hull, and combinations of pools form segments that delineate this convex hull. Therefore, the efficient frontier consists of combinations of two pools, with the selected pools depending on the level of variance \(\sigma^2\), as illustrated in \cref{ex:mean_variance_efficient_frontier}.

\begin{ex}\label{ex:mean_variance_efficient_frontier}
We consider the same miner and the same mining pools as in \cref{ex:mean_variance_risk_aversion}, but now illustrate the mean-variance efficient frontier in \cref{fig:efficient_frontier}. Depending on the level of the variance chosen, the miner directs her computing power to a combination of Pools $2$ and $3$, or a combination of Pools $0$ and $2$.

\begin{figure}[!ht]
    \includegraphics[width =0.6\textwidth]{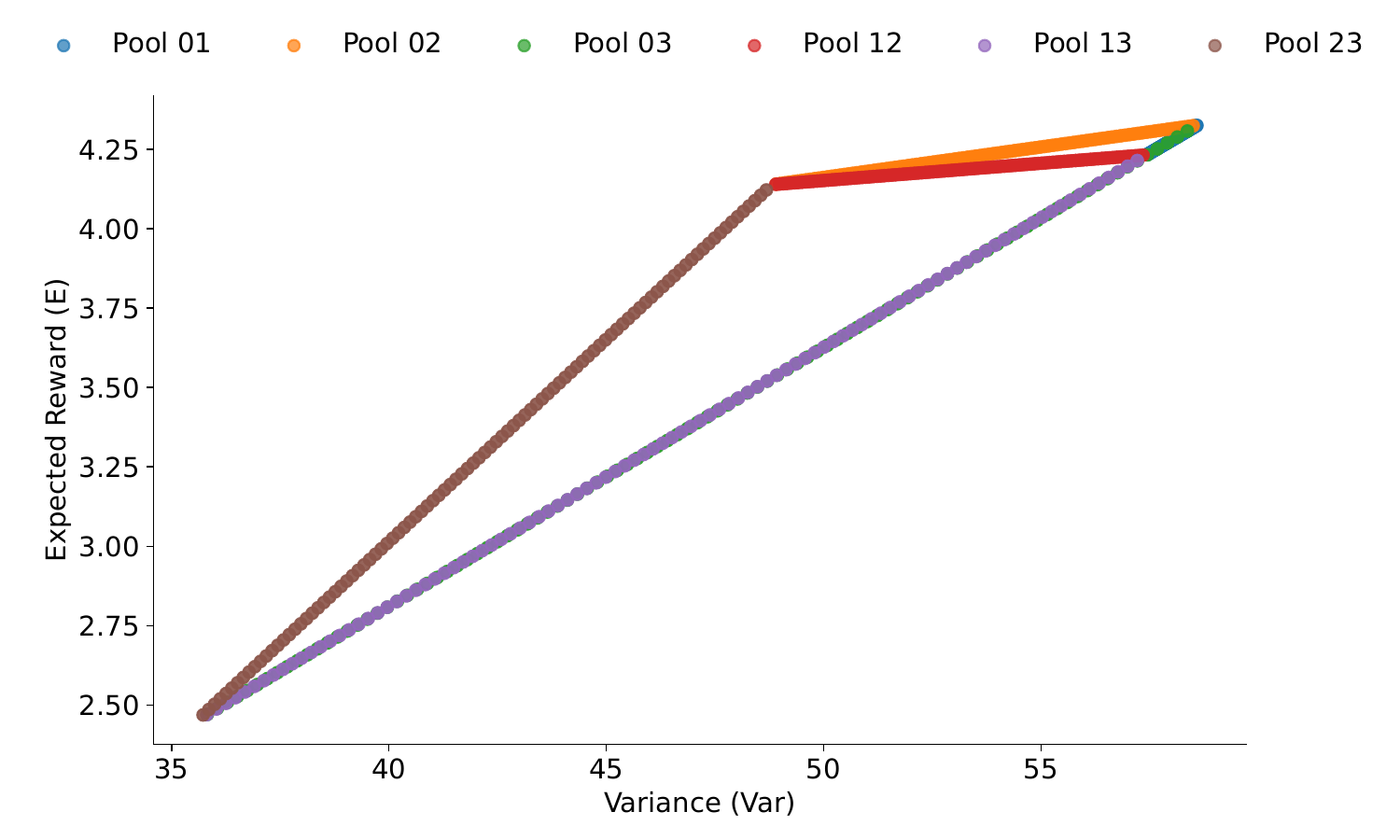}
    \centering
    \caption{Efficient frontier for $n = 4$ pools.}
    \label{fig:efficient_frontier}
\end{figure}
\end{ex}

\section{Expected discounted dividend maximization}\label{sec:optimal_revenue}
A standard performance indicator in insurance risk theory is the expected discounted dividends up to ruin. The process \((X_t)_{t \geq 0}\), defined in \eqref{eq:surplus_miner_split_accross_mining_pools}, represents the surplus of the miner. From this surplus, the miner extracts a portion according to a strategy \(\mathcal{S}\) as dividends. The dividend accumulation up to time \(t \geq 0\) is given by a non-decreasing process \(\left(L^{\mathcal{S}}_t\right)_{t \geq 0}\). To any revenue strategy \(\mathcal{S}\) is associated a controlled process
\[
U_t^{\mathcal{S}} = X_t - L_t^{\mathcal{S}}.
\]
The goal is to maximize the expected discounted income defined as:
\begin{equation}\label{eq:expected_discounted_mining_income}
V_{\mathcal{S}}(x) = \mathbb{E}_x\left[\int_{0}^{\sigma^\mathcal{S}} e^{-q t} \text{d}L_t^{\mathcal{S}}\right],
\end{equation}
where
\[
\sigma^\mathcal{S} = \inf\{t \geq 0 \text{ ; } U_t^\mathcal{S} < 0\},
\]
is the ruin time, \(\mathbb{E}_x(\cdot)\) is the expectation provided that \(X_0 = x\), and \(q \geq 0\) is the discount rate. The process \((X_t)_{t \geq 0}\) is a spectrally positive Lévy process; its Laplace exponent is defined by:
\[
\psi(\theta) = \log \mathbb{E}_0(e^{-\theta X_1}) = c\theta + \mu\left(\sum_{k=0}^n p_k e^{-b_k \theta} - 1\right).
\]
For such a wealth process, it is known that a barrier strategy is optimal, as shown in the work of \citet{Bayraktar2013}. Whenever the wealth of the process exceeds some level \(a\), the surplus in excess of \(a\) is cashed out immediately. The controlled process becomes the wealth process reflected at the level \(a\). The evaluation of the value function
\[
V_{\mathcal{S}}(x) := V(x, a),
\]
and the determination of the optimal level of the barrier
\[
a^\ast = \underset{a > 0}{\text{argmax}} V(x, a)
\]
follow from introducing the scale functions of \((X_t)_{t \geq 0}\). The \(q\)-scale function of \((X_t)_{t \geq 0}\) is defined through its Laplace transform as:
\begin{equation}\label{eq:Wq_function}
\int_{0}^{\infty} W^{(q)}(x) e^{-x \theta} \text{d}x = \frac{1}{\psi(\theta) - q}, \text{ } \theta < \phi(q),
\end{equation}
where
\begin{equation}\label{eq:phi}
\phi(q) = \sup\{\theta \geq 0 \text{ ; } \psi(\theta) = q\}.
\end{equation}
\begin{remark}\label{rem:ruin_proba}
The function $\phi$ is linked to the ruin probability of the miner as we have 
\begin{equation}\label{eq:ruin_proba_miner}
\varphi(x):=\Prob_x(\tau_0 < \infty) = \e^{-\phi(0)\cdot x},
\end{equation}
where $\tau_0 = \inf\{t\geq 0\text{ ;  }X_t \leq0\}$. We will use the ruin probability of a miner that does solo mining later to set the initial reserves. 
\end{remark}
We further define:
\begin{equation}\label{eq:Zq}
Z^{(q)}(x) = 1 + q \int_{0}^x W^{(q)}(y) \text{d}y,
\end{equation}
and
\begin{equation}\label{eq:Zq_bar}
\bar{Z}^{(q)}(x) = \int_{0}^x Z^{(q)}(y) \text{d}y.
\end{equation}
As we extend \(x \mapsto W^{(q)}\) to the entire real axis by setting \(W^{(q)}(x) = 0\) for \(x < 0\), we have:
\[
Z^{(q)}(x) = 1 \text{ and } \bar{Z}^{(q)}(x) = x, \text{ for } x \leq 0.
\]
It was shown in \citet[Lemma 2.1]{Bayraktar2013} (leveraging on \citet[Theorem 1]{Avram2007}) that the value function can be expressed as:
\begin{equation}\label{eq:value_function}
V(x; a) = -\kappa(a - x) + \frac{Z^{(q)}(a - x)}{Z^{(q)}(a)} \kappa(a), \text{ } x \leq a,
\end{equation}
where
\[
\kappa(y) = \bar{Z}^{(q)}(y) - \frac{Z^{(q)}(y)}{\phi(q)} - \frac{\psi'(0)}{q}.
\]
The optimal barrier level is given by:
\begin{equation}\label{eq:optimal_barrier}
a^{\ast} = \left(\bar{Z}^{(q)}\right)^{-1}\left(-\frac{\psi'(0)}{q}\right),
\end{equation}
which follows from differentiating \(V(x, a)\) with respect to \(a\). To each hashpower distribution \(w\) is associated an optimal level \(a^{\ast}_w\) and an expected mining income \(V(x; a^\ast_w)\). What we seek is the optimal hashpower distribution that maximizes the potential mining income such that:
\begin{equation}\label{eq:optimization_problem_dividend}
w^\ast = \underset{w \in \Delta^n}{\text{argmax}} V(x; a^\ast_w).
\end{equation}
The scale functions usually do not admit closed-form expressions, and we do not have any for the considered wealth process \eqref{eq:surplus_miner_split_accross_mining_pools}. The challenge is to find numerical methods to evaluate all these scale functions. The standard approach to calculating \(W^q\) is Laplace transform inversion. However, using numerical techniques followed by numerical integration will not be practical for searching the space of \(w\) to find the maximum of \(V(x; a^\ast_w)\). Fortunately, for the wealth process \((X_t)_{t \geq 0}\) the Laplace transform may be inverted analytically, resulting in formulas amenable for numerical evaluations, as shown in the following result.

\begin{prop}\label{prop:scale_function_formula}
Denote by $b^\ast = \underset{k = 0,\ldots, n}{\min} b_k.$
We have 
\begin{equation}\label{eq:Wq_formula}
W^{(q)}(x) = \frac{1}{c}\sum_{j = 0}^{\lfloor x/b^\ast\rfloor}\left(-\frac{\mu}{\mu+q}\right)^j\sum_{i_0+\ldots+ i_n = j}\binom{j}{i_0,\ldots, i_n}p_0^{i_{0}}\ldots p_n^{i_{n}}g\left[\frac{\mu+q}{c}\left(x-\sum_{k=0}^ni_kb_k\right), j\right],
\end{equation}
\begin{equation}\label{eq:Zq_formula}
Z^{q}(x)  = 1 + q\sum_{j = 0}^{\lfloor x/b^\ast\rfloor}\frac{(-\mu)^j}{(\mu+q)^{j+1}}\sum_{i_0+\ldots+ i_n = j}\binom{j}{i_0,\ldots, i_n}p_0^{i_{0}}\ldots p_n^{i_{n}}G\left[\frac{\mu+q}{c}\left(x-\sum_{k=0}^ni_kb_k\right), j\right],
\end{equation}
and 
\begin{equation}\label{eq:barZq_formula}
\bar{Z}^{q}(x) = x + q c\sum_{j = 0}^{\lfloor x/b^\ast\rfloor}\frac{(-\mu)^j}{(\mu+q)^{j+2}}\sum_{i_0+\ldots+ i_n = j}\binom{j}{i_0,\ldots, i_n}p_0^{i_{0}}\ldots p_n^{i_{n}}\bar G\left[\frac{\mu+q}{c}\left(x-\sum_{k=0}^ni_kb_k\right), j\right],
\end{equation}
where the functions $g, G$ and $\bar G$ are given by
$$
g(x,j) = \frac{e^{x}x^j}{j!}H(x),\text{ }G(x,j) = \int_0^x\frac{e^{y}y^j}{j!}\text{d}yH(x),\text{ and }\bar G(x,j) = \int_0^x\int_0^y\frac{e^{z}z^j}{j!}\text{d}z\text{d}yH(x),
$$
with $H(\cdot)$ the Heaviside step function.
\end{prop}
\begin{proof}
For the wealth process $(X_t)_{t\geq 0}$ defined in \eqref{eq:surplus_miner_split_accross_mining_pools}, we have 
\begin{eqnarray}
\int_{0}^{\infty}W^{(q)}(x)e^{-x\theta}\text{d}x &= &\frac{1}{c\theta +\mu\left(\sum_{k=0}^np_ke^{-b_k\theta} - 1\right) - q}\nonumber\\
&= &\frac{1}{c}\frac{1}{\theta - \frac{\mu+q}{c}}\frac{1}{1+\frac{\mu\sum_{k=0}^np_ke^{-b_k\theta}}{\theta - \frac{\mu+q}{c}}}\nonumber\\
&= &\frac{1}{c}\frac{1}{\theta - \frac{\mu+q}{c}}\sum_{j=0}^{\infty}\left(-\frac{\mu\sum_{k=0}^np_ke^{-b_k\theta}}{\theta - \frac{\mu+q}{c}}\right)^j\nonumber\\
&=&\frac{1}{c}\sum_{j=0}^{\infty}\frac{(-\mu)^j}{\left(\theta - \frac{\mu+q}{c}\right)^{j+1}}\sum_{i_0+\ldots+ i_n = j}\binom{j}{i_0,\ldots, i_n}p_0^{i_{0}}\ldots p_n^{i_{n}}e^{-\theta\sum_{k=0}^ni_kb_k}.\label{eq:proof1}
\end{eqnarray}
Note that 
$$
\frac{1}{\left(\theta - \frac{\mu + q}{c}\right)^{j+1}} = \int^{+\infty}_0e^{-\theta x}\frac{e^{\frac{\mu + q}{c}x}x^j}{j!}\text{d}x,
$$
which, when inserted in \eqref{eq:proof1}, yields
\begin{eqnarray*}
\int_{0}^{\infty}W^{(q)}(x)e^{-x\theta}\text{d}x &= &\int_0^{\infty}\frac{1}{c}\sum_{j=0}^{\infty}\frac{(-\mu)^j}{\left(\theta - \frac{\mu+q}{c}\right)^{j+1}}\sum_{i_0+\ldots+ i_n = j}\binom{j}{i_0,\ldots, i_n}p_0^{i_{0}}\ldots p_n^{i_{n}}\\
&\times&e^{-\theta\sum_{k=0}^ni_kb_k}g\left[\frac{\mu+q}{c}\left(x-\sum_{k=0}^ni_kb_k\right), j\right]e^{-\theta x}\text{d}x,
\end{eqnarray*}
after a change of variable. It holds that 
\begin{equation}\label{eq:proof2}
W^{(q)}(x) = \frac{1}{c}\sum_{j = 0}^{\infty}\left(-\frac{\mu}{\mu+q}\right)^j\sum_{i_0+\ldots+ i_n = j}\binom{j}{i_0,\ldots, i_n}p_0^{i_{0}}\ldots p_n^{i_{n}}g\left[\frac{\mu+q}{c}\left(x-\sum_{k=0}^ni_kb_k\right), j\right]
\end{equation}
through Laplace transform inversion. To conclude, note that $\sum_{k=0}^ni_kb_k > j\cdot b^\ast$ for all $i_0,\ldots, i_n$ such that $j = i_0+\ldots+i_n$. We can write
$$
W^{(q)}(x) = \frac{1}{c}\sum_{j = 0}^{\infty}\left(-\frac{\mu}{\mu+q}\right)^j\sum_{i_0+\ldots+ i_n = j}\binom{j}{i_0,\ldots, i_n}p_0^{i_{0}}\ldots p_n^{i_{n}}g\left[\frac{\mu+q}{c}\left(x-\sum_{k=0}^ni_kb_k\right), j\right]H(x- j\cdot b^\ast),
$$
and bound the sum in $j$ as in \eqref{eq:Wq_formula}. To get \eqref{eq:Zq}, we start again from \eqref{eq:proof2}, integration yields 
\begin{equation}\label{eq:proof3}
\int_0^x W^{(q)}(y)\text{dy} = \frac{1}{c}\sum_{j = 0}^{\infty}\left(-\frac{\mu}{\mu+q}\right)^j\sum_{i_0+\ldots+ i_n = j}\binom{j}{i_0,\ldots, i_n}p_0^{i_{0}}\ldots p_n^{i_{n}}\int_0^xg\left[\frac{\mu+q}{c}\left(y-\sum_{k=0}^ni_kb_k\right), j\right]\text{d}y.
\end{equation}
Notice that 
\begin{eqnarray*}
\int_0^xg\left[\frac{\mu+q}{c}\left(y-\sum_{k=0}^ni_kb_k\right), j\right]\text{d}y&=&\int_{\sum_{k=0}^ni_kb_k}^{x\vee \sum_{k=0}^ni_kb_k}\frac{e^{\frac{\mu+q}{c}\left(y-\sum_{k=0}^ni_kb_k\right)}\left[\frac{\mu+q}{c}\left(y-\sum_{k=0}^ni_kb_k\right)\right]^j}{j!}\text{d}y\\
&=&\frac{c}{\mu +q}\int_{0}^{\frac{\mu+q}{c}\left(x-\sum_{k=0}^ni_kb_k\right)\vee 0}\frac{e^{z}z^j}{j!}\text{d}z\\
&=&\frac{c}{\mu +q}H\left[\frac{\mu+q}{c}\left(x-\sum_{k=0}^ni_kb_k\right),j\right].
\end{eqnarray*}
Reinserting in \eqref{eq:proof3} then leads to \eqref{eq:Zq_formula}. The formula \eqref{eq:barZq_formula} follows from integrating \eqref{eq:Zq_formula} using the exact same algebraic trick. Similar computations can be found in the work of \citet{Albrecher2014}.
\end{proof}
We now have all the necessary components to evaluate the net present value of dividends \( V(x, a^\ast) \) at the optimal barrier level \( a^\ast \). The evaluations of \( G \) and \( \bar{G} \) follow from successive integration by parts as:
\[
G(x, j) = \begin{cases}
\sum_{i = 1}^{j} (-1)^i g(x, i) - 1, & \text{if } j \text{ is even}, \\
\sum_{i = 1}^{j} (-1)^{i+1} g(x, i) + 1, & \text{otherwise},
\end{cases}
\]
and
\[
\bar{G}(x, j) = \begin{cases}
\sum_{i = 1}^{j} (-1)^i G(x, i) - 1, & \text{if } j \text{ is even}, \\
\sum_{i = 1}^{j} (-1)^{i+1} G(x, i) + 1, & \text{otherwise}.
\end{cases}
\]
The function \( \phi(\cdot) \), on the other hand, requires the use of a root-finding algorithm, such as the bisection method. Through the following example, we see how the formulas simplify if the miner is solo mining or, alternatively, mining in only one pool.
\begin{remark}
If the miner is investing all her mining resources in solo mining then we have 
$$
\psi(\theta) = c\theta +\lambda\left(e^{-b\theta} - 1\right),
$$
and the function $q\mapsto \phi$ is given explicitly in terms of the $W$ Lambert function, see \citet{Corless1996}, as 
$$
\phi(q)= \frac{W\left[\frac{\lambda b}{c}e^{\frac{(q+\lambda)b}{c}} \right]}{b} + \frac{(q+\lambda)}{c}. 
$$ 
The formulas of \cref{prop:scale_function_formula} also simplify to 
\begin{equation*}
W^{(q)}(x) = \frac{1}{c}\sum_{j = 0}^{\lfloor x/b\rfloor}\left(-\frac{\lambda}{\lambda+q}\right)^jg\left[\frac{\lambda+q}{c}\left(x-j\cdot b\right), j\right],
\end{equation*}
\begin{equation*}
Z^{q}(x)  = 1 + q\sum_{j = 0}^{\lfloor x/b\rfloor}\frac{(-\lambda)^j}{(\lambda+q)^{j+1}}G\left[\frac{\lambda+q}{c}\left(x-j\cdot b\right), j\right],
\end{equation*}
and 
\begin{equation*}
\bar{Z}^{q}(x) = x + q c\sum_{j = 0}^{\lfloor x/b\rfloor}\frac{(-\lambda)^j}{(\lambda+q)^{j+2}}\bar G\left[\frac{\lambda+q}{c}\left(x-j\cdot b\right), j\right].
\end{equation*}
\end{remark}
The number of summands in the formulas of the scale functions in \cref{prop:scale_function_formula} increases rapidly as more mining pools are considered. The expression for the discounted expected dividend requires the use of numerical optimization routines to solve the problem \eqref{eq:optimization_problem_dividend}. We therefore adopt a bottom-up strategy: we start by calculating the expected discounted dividend within each pool, then look for the best combination of two pools, followed by three, and so on. If increasing the number of pools in the combinations does not improve the objective function, we stop the process. When considering combinations of two pools, we use a simple univariate root-finding algorithm. When the number of pools exceeds three, we employ a Particle Swarm Optimization (PSO) algorithm. PSO algorithms are well-suited for optimization problems where gradient information of the objective function is unavailable.
\begin{ex}\label{ex:dividend_maximization}
We consider the same miner and mining pools as in the previous examples. The intial wealth of the miner is set to 
\[
  x = \varphi^{-1}(0.5) = 3.939,
\]
where $\varphi$ is the ruin probability of the considered miner in the case where she decides to mine solo. The value functions for the case where the miner decides to limit herself to only one pool are provided in \cref{tab:expected_discounted_dividend_for_solo_mining_or_mining_in_one_pool}. Pool \(2\) offers the highest expected discounted dividend. Combining two pools does not lead to any improvement; consequently, the miner allocates all of her computing resources to Pool \(2\).
\begin{table}[!h] 
\centering
\begin{tabular}{l|rrrrr}
\toprule
\footnotesize
pool &  share discovery rate &  share reward &  pool fee &  difficulty reduction &     V \\
\midrule
\#2 &                 7.059 &         2.630 &     0.010 &                  0.85 & 6.027 \\
solo &                 6.000 &         3.125 &     0.000 &                  1.00 & 6.008 \\
  \#1 &                 6.061 &         3.078 &     0.005 &                  0.99 & 5.935 \\
  \#3 &                 8.000 &         2.109 &     0.100 &                  0.75 & 4.694 \\
\bottomrule
\end{tabular}
\caption{Expected discounted dividend for solo mining or mining in one pool.}
\label{tab:expected_discounted_dividend_for_solo_mining_or_mining_in_one_pool}
\end{table}
\end{ex}
The concept of expected discounted dividend is more challenging to grasp; however, it offers two significant advantages over the mean-variance approach. First, it eliminates the need to explicitly set the risk preferences of the agent, making it more objective. Second, variance tends to penalize large deviations from the mean, regardless of whether they are positive or negative. This can be particularly misleading in our scenario, where large jumps, although infrequent, can significantly enhance overall returns. By focusing solely on variance, we risk unduly penalizing these beneficial large jumps, thereby obscuring the true potential of the investment strategy. Instead, a risk measure that differentiates between the nature and impact of these jumps, such as the probability of ruin, provides a more accurate assessment of the profit and liability. The expected discounted dividend also accounts for the initial wealth and does not require the specification of a time horizon (although we noted that the mean-variance approach was actually agnostic to the choice of a time horizon).

\begin{color}{blue}\end{color}
\section{Numerical illustrations and decentralization of mining network}\label{sec:numerical_illustrations}
\begin{figure}[!ht]
  \begin{center}
    \subfloat[Mean-Variance utility]{
      \includegraphics[width=0.5\textwidth]{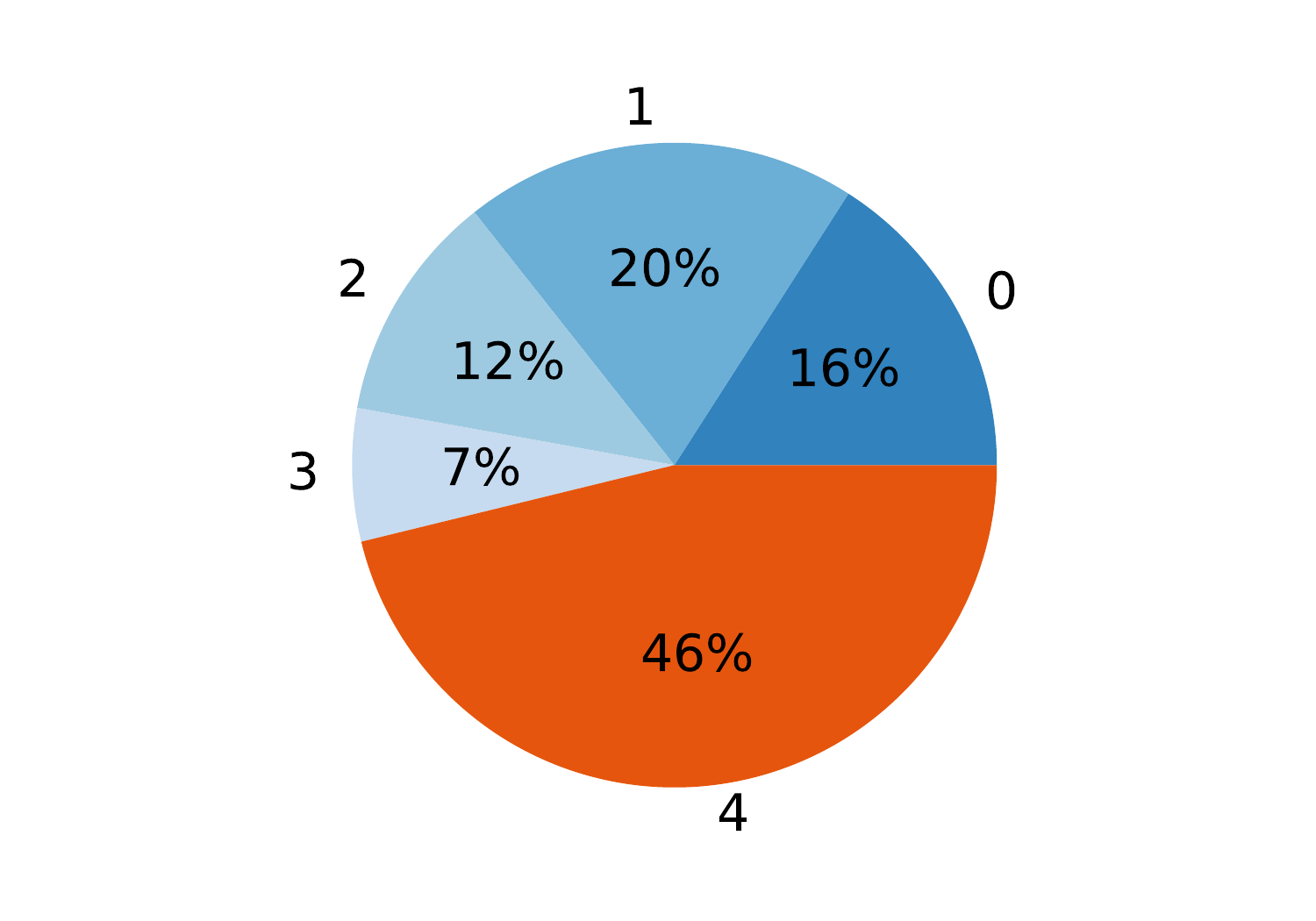}
      \label{sub:mean_variance_utility}
    }
    \subfloat[Mean-Variance Efficient frontier]{
      \includegraphics[width=0.5\textwidth]{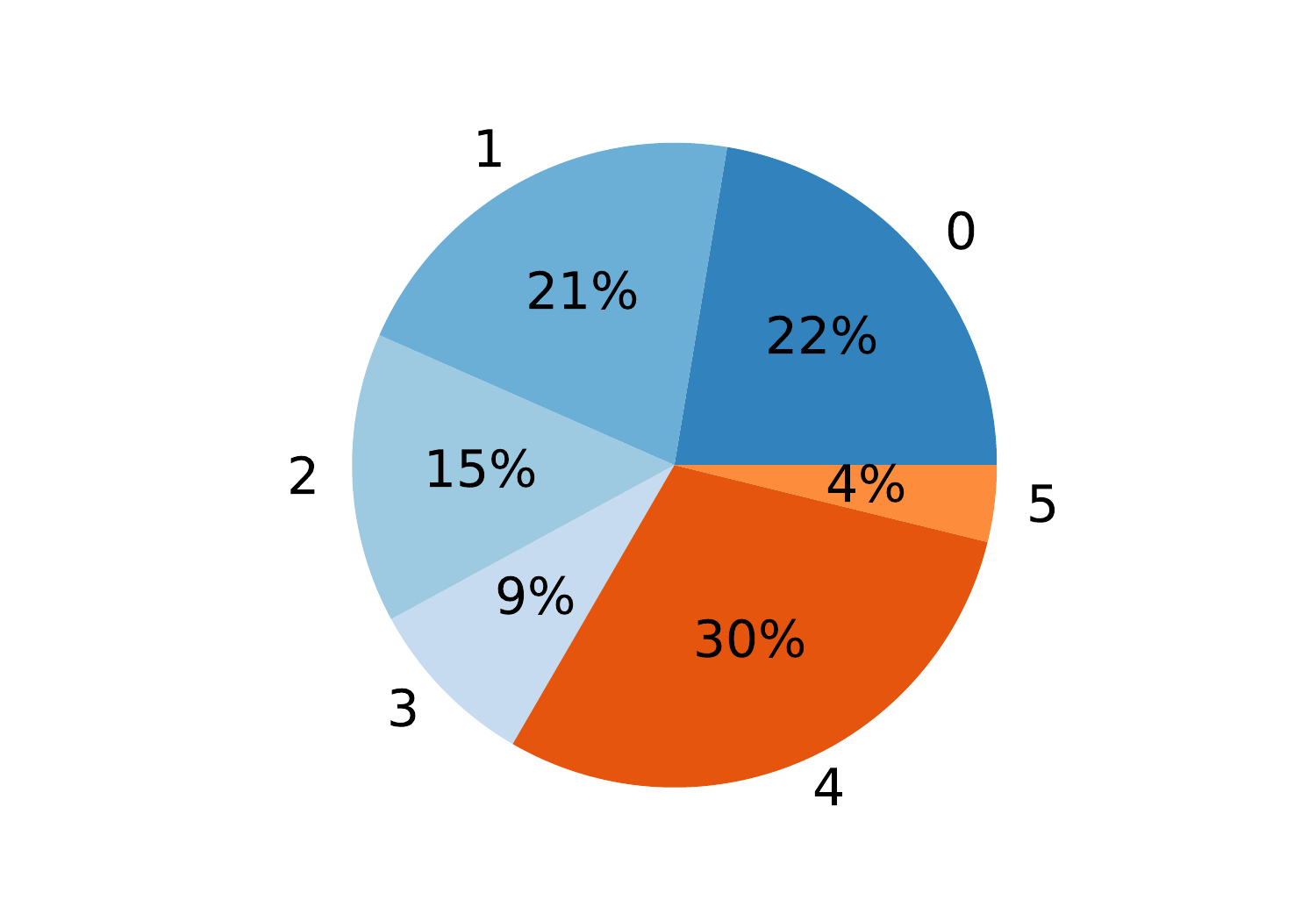}
      \label{sub:mean_variance_frontier}
    }\\
    \subfloat[Expected discounted dividends]{
      \includegraphics[width=0.5\textwidth]{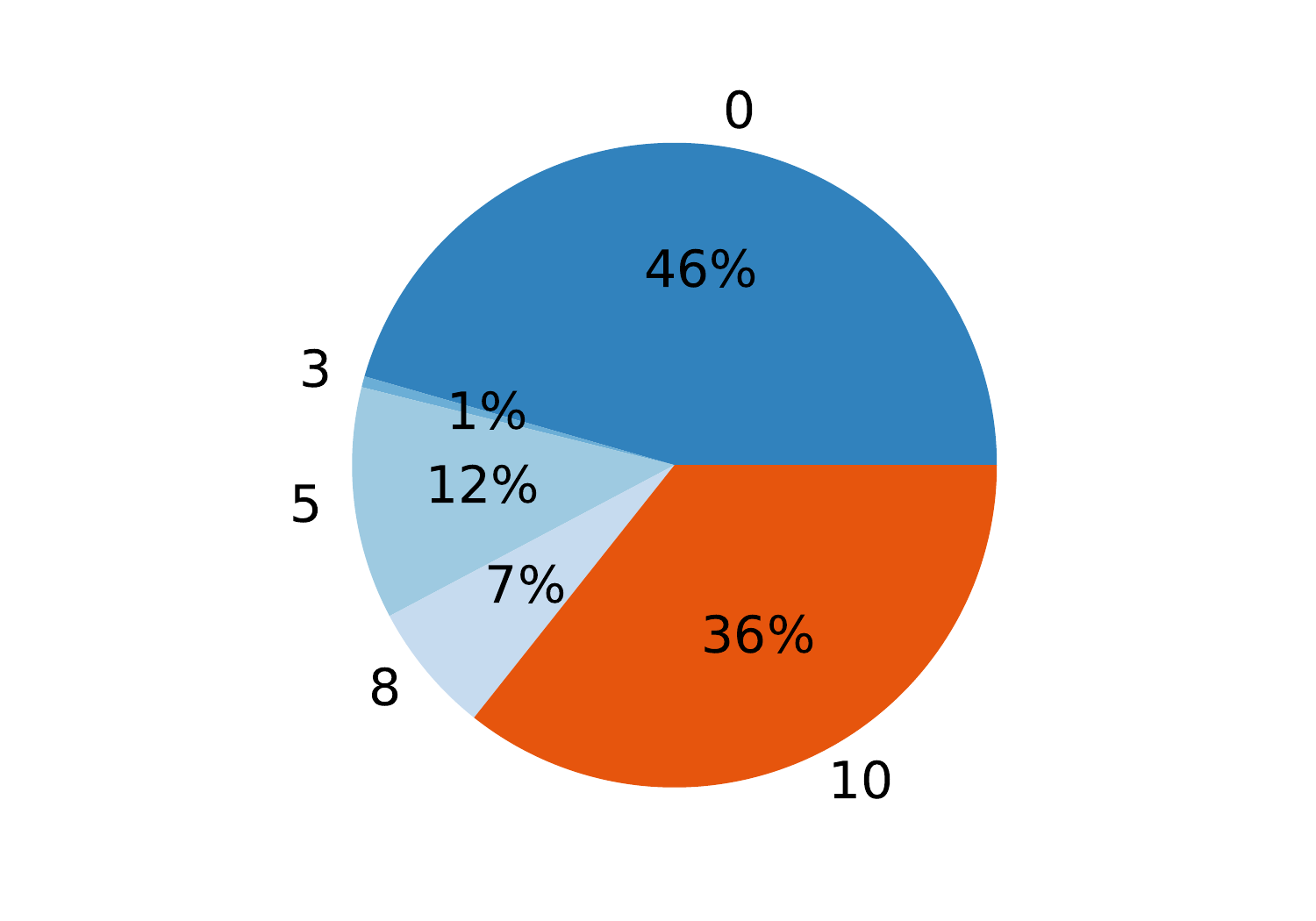}
      \label{sub:dividend}
    }
    \caption{Hashpower distribution among mining pools depending on the selected criterion.}
    \label{fig:pie_hashpower}
  \end{center}
\end{figure}

A simulation study is conducted to explore the decentralization of a blockchain network composed of miners and mining pools. We investigate the distribution of the computing power of miners with characteristics \((\lambda_l, c_l)\) for \(l = 1, \ldots, m\) among mining pools with characteristics \((\delta_k, f_k)\) for \(k = 0, \ldots, n\). The network's hashrate is \(\lambda = 6\) (6 blocks per hour on average). We distribute the hashrate to the miners as
\[
\left(\begin{array}{ccc}\lambda_1 & \ldots & \lambda_m\end{array}\right) \sim \text{Dir}(1, \ldots, 1) \cdot \lambda.
\]
The operational cost of each miner is given by
\[
c_l = \frac{\lambda_l \cdot b}{(1 + \eta_l)}, \text{ for } l = 1, \ldots, n,
\]
where \(\eta_l \overset{i.i.d.}{\sim} \text{Unif}([0.04, 0.1])\). We set the initial wealth to achieve a specific risk level characterized through the ruin probability as
\[
\varphi_l(x_l) = \beta_l, \text{ for } l = 1, \ldots, m,
\]
where \(\beta_l \overset{i.i.d.}{\sim} \text{Beta}(1.5, 1.5)\) and \(\varphi_l\) is the ruin probability for miner \(l\) when mining solo. The mean-variance trade-off requires selecting a risk aversion parameter that we sample as
\[
\gamma_l \sim \text{Unif}(0, 1), \text{ for } l = 1, \ldots, m.
\]
The miners can invest their computing resources in \(n\) mining pools with characteristics \((f_1, \delta_1), \ldots, (f_n, \delta_n)\). The miners can also mine solo, which is equivalent to mining for mining pool \(0\) with characteristics \(f_0 = 0\) and \(\delta_0 = 1\). We set the pool fees as:
\[
f_k \sim \text{Unif}([0, 0.04]) \text{ for } k = 1, \ldots, n.
\]
The difficulty reduction ensures that the expected discounted dividend of an average miner with characteristics
\[
\bar{\lambda} = \frac{\lambda}{m}, \quad \bar{c} = \frac{\bar{\lambda} \cdot b}{(1 + 0.17)}, \quad \text{and} \quad \bar{x} = \bar{\psi}^{-1}(0.5)
\]
in the network is the same for all the mining pools. We set \(n = 10\) and \(m = 25\) to obtain the hashpower distribution illustrated in \cref{fig:pie_hashpower}.

\section{Conclusion}\label{sec:conclusion}

We have proposed two types of performance indicators to help blockchain miners achieve a trade-off between profitability and risk. These indicators can assist miners in selecting Pay-Per-Share (PPS) mining pools toward which to direct their computing resources. This approach provides insights into how the network hashpower should theoretically be distributed across mining pools and whether a few mining pools concentrate large portions of the total hashpower. The theoretical distribution can be compared to the observed distribution, which can be retrieved from various online sources of varying reliability, such as \url{https://miningpoolstats.stream/}.\\

\noindent Mean-variance based indicators are easy to understand and quick to compute. However, the arbitrary choice of a risk aversion parameter and the penalization of large upward jumps are the main drawbacks of such indicators. The expected discounted dividend, while more involved to calculate, integrates risk in a more sophisticated manner and does not require the selection of a risk aversion parameter.\\

\noindent There remains ample room for future work. The analysis could include other reward systems, such as Pay-Per-Last-N-Shares (PPLNS). The capital gains component in the wealth process is not stationary, as both the reward arrival and the portion of the reward earned by the miner depend on the size of the pool, which is a moving target. The inclusion of transaction fees in the wealth process is an interesting direction that would introduce non-stationarity to the wealth process, as the level of transaction fees is driven by the congestion of the blockchain network, i.e., how many transactions are broadcast to the network. A comparison between the Pay-Per-Share Plus (PPS+) and Full Pay-Per-Share (FPPS) reward systems (recall \cref{ssec:mining_pool}) would be appropriate. Another direction could be the inclusion of the exchange rate of fiat against cryptocurrencies. The calculation of the expected discounted dividend could be enhanced by leveraging previous work from the actuarial science literature, specifically the study of the dividend problem when payments are made in a foreign currency, see e.g.\ \citet{Eisenberg2016}. Additionally, mining pools offer the possibility of mining different cryptocurrencies. In this context, the mean-variance trade-off would be a natural solution to include the expected return and volatility of the different crypto assets.\\

\noindent In forthcoming research, we plan to address the pool manager's problem, specifically how a pool manager can adjust the Pay-Per-Share (PPS) parameters to attract more miners while optimizing her own profit and risk trade-off. The wealth process of a PPS pool manager is a two-sided jump process, with upward jumps corresponding to block rewards and downward jumps corresponding to share payments; however, the dividend problem for two-sided jump processes is not yet exhaustively understood in the current literature.
We are also planning to introduce a more dynamic aspect to the miner and pool manager's problem, where the mining power and mining pool parameters can be adjusted in real time.
  
\section*{Acknowledgmements} 
Pierre-O. Goffard's work is supported  the ANR project
BLOCKFI\footnote{\url{https://anr.fr/Project-ANR-24-CE38-7885}}. All the plots and tables of the paper can be reproduced using the python code available from the following github repository \url{https://github.com/LaGauffre/optimal_blockchain_mining}.

\end{document}